\documentclass[runningheads]{llncs}
\usepackage{comment}
\usepackage{tabularx} 
\usepackage{enumitem}
\usepackage{booktabs}
\usepackage{multirow}
\usepackage[table]{xcolor}
\usepackage{algorithm}
\usepackage{algpseudocode}
\usepackage{amsmath}
\DeclareMathOperator*{\argmax}{arg\,max}
\usepackage{hyperref}
\hypersetup{
  colorlinks=true,
  linkcolor=blue,    
  citecolor=blue,    
  urlcolor=blue      
}
\usepackage[T1]{fontenc}
\usepackage{graphicx}

\usepackage{soul}      
\usepackage{xcolor}    
\sethlcolor{yellow}    
\begin{document}
\title{Dataset-Aware Cold-Start Active Learning for Annotation-Efficient 3D Medical Image Segmentation}
\titlerunning{CSCS for 3D medical image segmentation}
%
\author{
Rémi Hattat\inst{1}
\and Marine Beaumont\inst{1,2}
\and Charline Bertholdt\inst{1,3}
\and Gabriela Hossu\inst{1,2}
\and Olivier Morel\inst{1,3}
\and Bailiang Chen\inst{1,2}
}
\authorrunning{Hattat et al.}
%
\institute{IADI (U1254), Inserm and Université de Lorraine, Nancy, France 
\and CIC-IT 1433, Inserm, Université de Lorraine, and CHRU Nancy, Nancy, France 
\and Université de Lorraine, CHRU-Nancy, Pôle de la femme, Nancy, France 
\\
\email{remi.hattat@univ-lorraine.fr}}
\maketitle              
\begin{abstract}

Deep learning for 3D medical image segmentation requires extensive
manual annotations, a bottleneck that active learning (AL) seeks
to alleviate by selecting the most informative samples for labeling.
However, most AL methods assume access to an initial labeled set,
leaving the cold-start problem largely unresolved: how to select
volumes from a fully unlabeled pool before any model is trained.
Existing cold-start strategies apply fixed selection criteria
regardless of dataset embedding structure or annotation budget, yet the optimal
balance between representative and difficult samples varies across
clinical tasks. We propose \textbf{CSCS} (\textbf{C}urriculum-\textbf{S}tratified \textbf{C}old-\textbf{S}tart),
a dataset-aware cold-start framework that models this trade-off using
self-supervised signals available prior to annotation. Each unlabeled
volume is scored through a weighted geometric mean of
typicality and reconstruction-based uncertainty, where the weighting
is determined by a closed-form pacing rule driven by two pool-level
statistics: the effective annotation budget and the
Difficulty-Coverage Ratio (DCR), a rank-based measure of the alignment
between difficulty and representativeness in the unlabeled pool. We evaluate CSCS on four 3D benchmarks
(BraTS, FeTA, Spleen, and an in-house fetal-MRI dataset), comparing against five baselines using nnU-Net for
downstream segmentation across multiple annotation regimes. Across datasets and budgets, CSCS shows consistently competitive performance, with the most pronounced gains
in low-to-mid annotation regimes. Overall, our results suggest that dataset-aware cold-start
initialization can improve the robustness of active learning for 3D
medical image segmentation by adapting selection to the structure of
the unlabeled pool in a learned embedding space. 

\end{abstract}

\section{Introduction}
\label{sec:intro}

Automated 3D medical image segmentation is increasingly used to support
quantitative analysis, anatomical assessment, and treatment planning
across imaging modalities, and deep learning has substantially improved
its accuracy~\cite{isensee2021nnunet,hatamizadeh2022swinunetr}. In
clinical practice, reliable segmentation underpins tasks such as
volumetric quantification, longitudinal follow-up, and surgical or
radiotherapy planning, where it also reduces the time burden and the
inter-expert variability of manual delineation. Yet these gains depend
critically on large-scale expert annotation, a prohibitive bottleneck in
volumetric imaging: manually labeling a single 3D scan requires tens of
minutes of specialist effort~\cite{tajbakhsh2020embracing,wang2021annotation}.
Active Learning (AL) mitigates this burden by iteratively selecting the
most informative samples for
annotation~\cite{settles2009active,wang2024survey}, but a practical
question precedes it: how to build a first reliable 3D segmentation model
when expert annotation is scarce and costly, and no labels yet exist to
guide the choice.

This is the cold-start problem. Warm-start AL
methods~\cite{gal2017deep,sener2018active,ash2020badge} assume an initial
labeled set $\mathcal{L}_0$ is already available before iterative
querying begins; cold-start AL concerns the preceding stage, where
$\mathcal{L}_0$ must be selected from a fully unlabeled pool before any
task-specific model can be trained. In practice this stage is often
handled by random initialization, which is unstable and whose
variability propagates into downstream
performance~\cite{chandra2021initial,nath2022warm}. Constructing an
informative $\mathcal{L}_0$ in a single selection step therefore remains
a central problem for 3D medical
segmentation~\cite{liu2023colossal,chen2023hacon}. Existing cold-start
approaches attack it from two complementary angles: diversity-based
methods use self-supervised representations to cover the unlabeled
distribution~\cite{yehuda2022probcover,hacohen2022budget,yuan2024foundation},
while uncertainty-based strategies prioritize difficult samples but are
hard to apply before a task-specific model
exists~\cite{gal2017deep,settles2009active}. The recent hybrid CSAL-3D~\cite{zhu2025csal3d}
combines the two by selecting uncertain samples within
diversity-preserving clusters through a fixed hierarchical scheme, and
related work has shown the strong influence of representation quality on
cold-start performance~\cite{levy2026coldstart,zhu2025medcalbench}.
Rigorous benchmarks for iterative 3D biomedical
AL~\cite{luth2025nnactive,traub2026clasp} and adaptive curriculum
strategies for medical image classification~\cite{ma2024acal} have also
appeared, though neither targets cold-start 3D segmentation. Across these
methods, selection relies on a fixed criterion, overlooking that
the optimal balance between representativeness and difficulty depends on
both the structure of the unlabeled pool and the annotation
budget~\cite{liu2023colossal}.

This dataset- and budget-dependence is the gap we address. Prior
budget-aware AL shows that opposite strategies can be preferable in low-
and high-budget regimes, and more generally that the best acquisition
strategy is problem-dependent~\cite{hacohen2022budget,hacohen2023select}:
representative samples give stable coverage at low budgets, while harder
or more uncertain samples become beneficial as the budget grows. In the
cold-start setting the dependency is sharper still, because the absence
of labels means difficulty estimated from self-supervised signals may
reflect either informative anatomical variability or uninformative
artifacts, and whether difficult samples are central or peripheral in the
embedding space is itself a dataset-level property. A single fixed rule
cannot capture this variability, which motivates a cold-start strategy
that adapts to the data it is given.

We propose \textbf{CSCS} (\textbf{C}urriculum-\textbf{S}tratified
\textbf{C}old-\textbf{S}tart), a dataset-aware cold-start framework for selecting
an informative initial annotation set before segmentation training
(Fig.~\ref{fig:pipeline}). CSCS does not introduce a new fixed query
criterion; instead, it adapts the balance between selection signals from
statistics of the unlabeled pool. From a pre-trained self-supervised
encoder it extracts two label-free signals: typicality $T(x)$, measuring
local density, and reconstruction-based uncertainty $U(x)$, a difficulty
proxy following CSAL-3D~\cite{zhu2025csal3d} and supported by
reconstruction-based performance prediction for
segmentation~\cite{bar2022performance}. Each candidate is scored as
$S(x)=T(x)^{1-\gamma}U(x)^{\gamma}$, where the pacing parameter
$\gamma\in[0,1]$ shifts selection from representativeness-first
($\gamma{=}0$) to difficulty-first ($\gamma{=}1$). Rather than fixing
$\gamma$, CSCS sets it through a closed-form rule based on two pool-level
statistics available before annotation: the annotation budget, summarized by an effective budget term, and the
Difficulty-Coverage Ratio (DCR), a rank correlation measuring whether
difficult samples tend to lie in dense or peripheral regions of the
embedding space. Cold-start selection is thereby adapted to both the
budget and the structure of the unlabeled pool.

\noindent\textbf{Contributions.}
\begin{enumerate}[noitemsep, topsep=2pt, leftmargin=*]
    \item We frame cold-start selection as a dataset-dependent
          trade-off between representativeness and difficulty, and
          introduce the DCR as a pool-level descriptor to characterize
          this trade-off without labels.
    \item We propose CSCS, a dataset-aware cold-start framework that
      adapts initial data selection through a closed-form pacing rule
      without dataset-specific tuning.
    \item We demonstrate that CSCS achieves the most stable performance
          across datasets and budgets, highlighting robustness rather
          than isolated dataset-specific gains.
\end{enumerate}

\noindent
Unlike fixed strategies, CSCS explicitly leverages dataset statistics to
adapt acquisition, framing cold-start initialization as a one-shot,
distribution-aware selection strategy rather than a fixed selection rule.

\section{Method}
\label{sec:method}

\subsection{Problem Formulation and Overview}

Cold-start active learning (CSAL) aims to construct the initial labeled
set $\mathcal{L}_0$ of size $B$ by selecting samples from a fully
unlabeled pool $\mathcal{U} = \{X_1, \ldots, X_N\}$ in a single
selection step, without access to any trained task-specific model or
prior annotations. The selected samples are then annotated by experts
and form $\mathcal{L}_0$, which is used to train the first segmentation
model. The quality of this selection
critically determines the effectiveness of the subsequent AL
pipeline~\cite{liu2023colossal,chandra2021initial}. Because no
corrective feedback is available before training, poor choices made at
this stage directly propagate to all later acquisition rounds.

\begin{figure}[t]
\centering
\includegraphics[width=\linewidth]{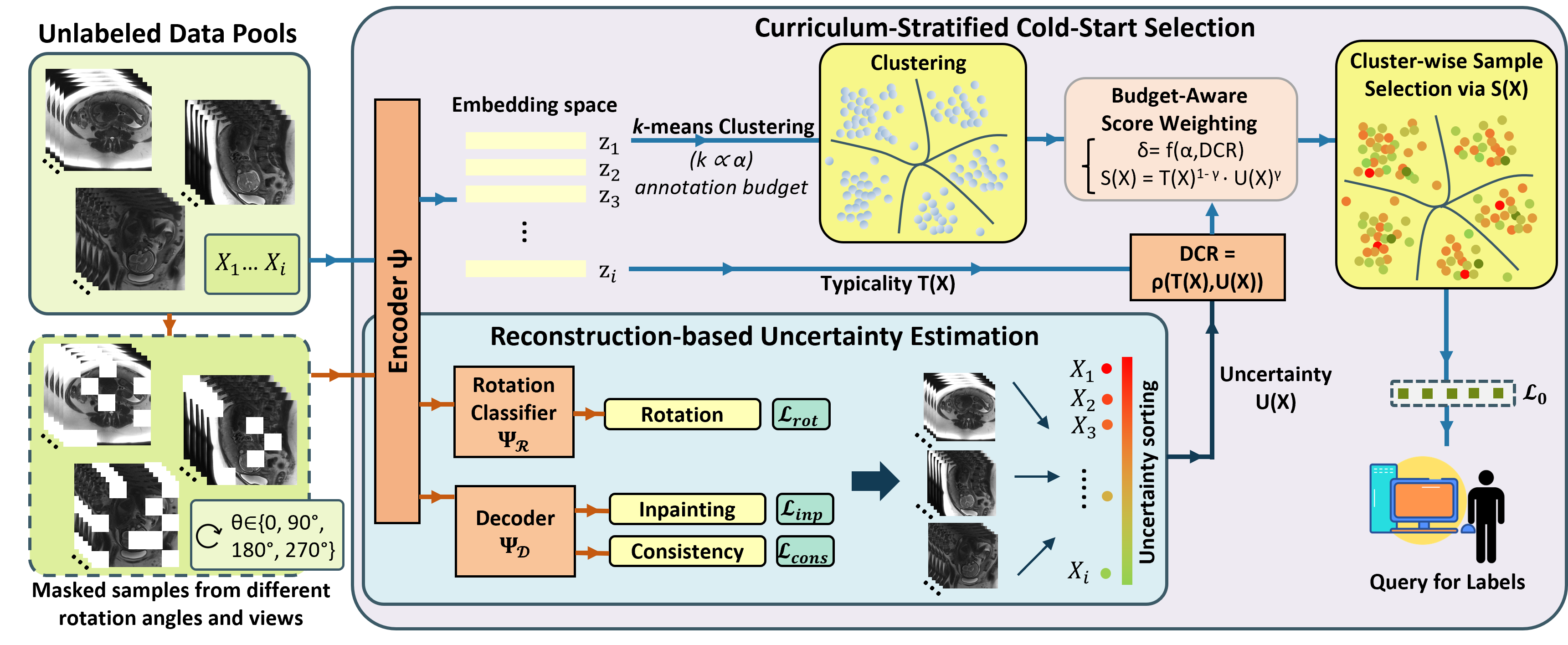}
\caption{Overview of the proposed CSCS pipeline. Self-supervised
learning (SSL) pretraining is performed on the unlabeled pool to extract,
for each volume $X_i$, an embedding $\mathbf{z}_i$, a typicality score
$T(X_i)$, and a reconstruction-based uncertainty score $U(X_i)$. After
diversity-preserving clustering into $B$ groups, one sample per cluster
is selected by maximizing the composite score
$S(X_i)=T(X_i)^{1-\gamma}\cdot U(X_i)^{\gamma}$. The pacing parameter
$\gamma$ is set via a closed-form rule from the effective budget and
the DCR, continuously interpolating between representativeness-first
and difficulty-first selection.}
\label{fig:pipeline}
\end{figure}

CSCS addresses this problem through a three-stage pipeline
(Fig.~\ref{fig:pipeline}). A self-supervised learning (SSL) pretraining stage first
extracts, for each volume, an embedding $\mathbf{z}_i$, an uncertainty
score $U(X_i)$, and a typicality score $T(X_i)$, together with a
pool-level statistic, the Difficulty-Coverage Ratio (DCR), that
characterizes whether difficult samples tend to lie in dense or
peripheral regions of the embedding space. Diversity-preserving
clustering then partitions the pool into $B$ groups. Finally, one
sample is selected per cluster by maximizing a composite score
$S(X_i) = T(X_i)^{1-\gamma} \cdot U(X_i)^{\gamma}$, where the pacing
parameter $\gamma$ controls the balance between representativeness and
difficulty and is determined from two pool-level quantities: the DCR and
an effective budget term. CSCS therefore separates pool-level adaptation from local sample selection: DCR and the effective budget determine how strongly the score
should favor typical, representative samples versus difficult, uncertain
samples, while cluster-wise maximization of $S$ selects one candidate per
region of the embedding space.

\subsection{SSL Pretraining and Pool Characterization}
\label{sec:ssl}

We adopt the masked inpainting SSL framework of Tang et
al.~\cite{tang2022swinunetr}, using a Swin-UNETR encoder $\Phi$
pretrained on $\mathcal{U}$ via multi-view reconstruction. This stage
is not a contribution of this work; rather, it provides a label-free
representation of the unlabeled pool, following the cold-start
formulation of CSAL-3D~\cite{zhu2025csal3d}. In CSCS, the frozen SSL
encoder is used to derive two complementary sample-level signals:
reconstruction-based uncertainty and local typicality.

\textbf{Uncertainty.} Uncertainty is widely used in deep learning as a
proxy for sample informativeness or difficulty~\cite{kendall2017uncertainties}.
In the cold-start setting, no task-specific segmentation model is
available to estimate predictive uncertainty. We therefore derive a
label-free uncertainty score from the SSL reconstruction task, following
the reconstruction-variance principle used in CSAL-3D~\cite{zhu2025csal3d}.
For each volume $X_i$, the encoder-decoder model produces $R$
reconstructions $\{\hat{X}_i^{(r)}\}_{r=1}^{R}$ from masked inputs under
different geometric views. The voxel-wise reconstruction variance is
aggregated into a scalar score:
\begin{equation}
    U(X_i) =
    \frac{1}{|\Omega_i|}
    \sum_{v \in \Omega_i}
    \mathrm{Var}_{r=1,\ldots,R}
    \left(\hat{X}_i^{(r)}(v)\right),
    \label{eq:uncertainty}
\end{equation}
where $\Omega_i$ denotes the set of reconstructed voxel positions.
Higher $U(X_i)$ indicates stronger disagreement between reconstructed
views and is used as a label-free proxy for sample difficulty.
The use of reconstruction-based signals as difficulty proxies is further
motivated by prior work showing that self-supervised reconstruction
quality can be informative for downstream segmentation
performance~\cite{bar2022performance}.

\textbf{Typicality.} We measure sample-level typicality as a local
density estimate in the embedding space, using the inverse average
distance to the $k$ nearest neighbors~\cite{hacohen2022budget}:
\begin{equation}
    T(X_i) = \left(\frac{1}{k} \sum_{X_j \in \text{KNN}_k(X_i)}
    \|\mathbf{z}_i - \mathbf{z}_j\|_2\right)^{-1}
    \label{eq:typicality}
\end{equation}
where  $\mathbf{z}_i$ denotes the embedding of volume
$X_i$, for $i=1,\ldots,N$. We set $k{=}20$ following prior
work~\cite{hacohen2022budget}, which provides a local density estimate
without relying on labels. High typicality indicates that a sample lies
near many neighboring samples in the learned representation space,
whereas low typicality indicates a more isolated or peripheral sample.
Both $U$ and $T$ are then min-max normalized to $[0,1]$ across the full
pool so that the two signals are comparable in the composite score and
the trade-off induced by $\gamma$ is not dominated by arbitrary scale
differences.

\textbf{Difficulty-Coverage Ratio.} Budget-aware AL
theory~\cite{hacohen2022budget} shows that the optimal transition
between easy-first and hard-first regimes depends on the structural
separation between easy and hard sub-populations. We capture this
property directly from SSL outputs through the Difficulty-Coverage
Ratio:
\begin{equation}
    \text{DCR} = \rho_s\!\left(\{U(X_i)\}_{i=1}^{N},\;
    \{T(X_i)\}_{i=1}^{N}\right)
    \label{eq:dcr}
\end{equation}
where $\rho_s$ is the Spearman rank
correlation~\cite{conover1999practical}. Spearman correlation is used instead of Pearson correlation because the
relationship between uncertainty and typicality is not expected to be
linear. By operating on ranks, Spearman correlation reduces the influence
of absolute scale differences between $U$ and $T$ and is invariant to
monotonic transformations of both quantities. DCR therefore reflects the
relative ordering between reconstruction-based difficulty and local
typicality, rather than their raw numerical values.

DCR describes pool geometry. It does not assess the absolute
reliability of the uncertainty signal: it
indicates whether difficult samples tend to lie in structurally central
or peripheral regions of the current pool. DCR does not rank samples
directly; it only calibrates the global trade-off parameter $\gamma$.
A positive DCR indicates that difficult samples tend to lie in dense,
representative regions, suggesting that uncertainty can be emphasized
without drifting toward outliers. A negative DCR reflects the opposite
situation: difficult samples are more likely to be peripheral, so
uncertainty becomes a riskier signal and a more conservative,
typicality-driven behavior is preferable. DCR is computed on the full
pool rather than per-cluster, because clusters in the cold-start regime
typically contain too few samples for a stable rank-correlation
estimate. In small-pool settings, pool-level estimates can still
exhibit non-negligible sampling variance; this further supports keeping
the influence of DCR moderate through the budget-dependent factor
introduced in Eq.~\ref{eq:gamma}. This is consistent with the
observation in~\cite{hacohen2022budget} that the transition between
easy- and hard-sample preference depends on the structural
organization of the data.

\subsection{Diversity-Preserving Clustering}
\label{sec:clustering}

Clustering enforces structural diversity by distributing selections
across the embedding space before applying score-based refinement.
We apply $k$-means++ clustering~\cite{arthur2007kmeans} with the number of clusters set to $B$
on the normalized embeddings $\{\mathbf{z}_i\}$. Setting the cluster
count equal to the annotation budget allows, in the ideal case, one
primary selection per cluster without additional diversity
constraints~\cite{sener2018active}. In this sense, clustering specifies
where samples may be selected, while the score defined below
specifies which sample is preferred within each region.

Because very small clusters provide unstable within-cluster rankings, we
merge clusters containing fewer than $s_{\min}=3$ samples before final
selection. Each small cluster is merged with its nearest neighboring cluster, where
cluster proximity is defined by the Euclidean distance between cluster
centroids in the normalized embedding space. If merging reduces the
number of resulting clusters below $B$, the remaining selection slots
are filled by global ordering of the unselected samples according to the
same composite score.

\subsection{Closed-Form Curriculum Pacing}
\label{sec:pacing}

\textbf{Composite score.}
Within each cluster $C_m$, we combine typicality and uncertainty
through a weighted geometric mean:
\begin{equation}
    S(X_i) = T(X_i)^{1-\gamma} \cdot U(X_i)^{\gamma},
    \qquad X_m^* = \argmax_{X_i \in C_m} S(X_i)
    \label{eq:score}
\end{equation}
where $X_m^*$ denotes the selected candidate from cluster $C_m$.
The initial query set is then formed by collecting one selected
candidate from each cluster,
$\mathcal{L}_0=\{X_m^*\}_{m=1}^{M}$, where $M$ is the number of clusters
after possible merging. If $M<B$, the remaining slots are filled by
globally ranking the unselected samples according to $S$ until
$|\mathcal{L}_0|=B$.

This formulation treats the two signals as complementary rather
than substitutable. The parameter $\gamma$ controls their relative
influence: increasing $\gamma$ gives more weight to uncertainty, whereas
decreasing $\gamma$ gives more weight to typicality. Moreover,
$S(X_i)=0$ whenever either term vanishes, preventing a sample from being
selected on the strength of a single criterion alone. The exponents sum
to one, keeping $S$ on the same scale as $T$ and $U$ regardless of
$\gamma$. At $\gamma{=}0.5$, Eq.~\ref{eq:score} reduces to the
unweighted geometric mean $\sqrt{T \cdot U}$; at the conceptual
extremes, $\gamma{=}0$ reduces to pure typicality ranking and
$\gamma{=}1$ to pure uncertainty ranking. These two extremes are used
only as ablation references and are not produced by the adaptive pacing
rule used in CSCS to set $\gamma$.

\textbf{Effective budget.}
The pacing rule combines two signals: the DCR, which encodes the
direction of the trade-off, and a budget-dependent term that controls
how strongly DCR should influence selection. When only a small number
of samples is selected, mistakes are costly and the curriculum should
remain conservative; larger budgets allow stronger adaptation to
dataset structure \cite{hacohen2022budget,hacohen2023select}.

We therefore introduce a confidence factor that must increase with
budget, remain bounded, and vary smoothly. The monotone saturating
transform $x/(1+x)$, mapping any positive input to $(0,1)$,
satisfies all three requirements.

As input to this transform, we require a budget descriptor that
captures both absolute count and relative coverage. These two
desiderata correspond to the extremes of a one-parameter family:
\begin{equation}
    \alpha_\beta = \frac{B}{N^\beta}, \quad \beta \in [0,1]
    \label{eq:alpha_family}
\end{equation}
where $\beta{=}0$ reduces to a count-only descriptor ($B$, ignoring
pool size) and $\beta{=}1$ to the ratio-only descriptor ($B/N$,
ignoring the absolute scale of the budget). We set $\beta{=}1/2$,
the parameter-free midpoint of this family, yielding
\begin{equation}
    \alpha_{\text{eff}} = \frac{B}{\sqrt{N}}
    \label{eq:alpha_eff}
\end{equation}
Importantly, $\alpha_{\text{eff}}$ is an internal quantity used
solely to modulate the pacing rule; it should not be confused with
the experimental ratio $\alpha = B/N$ used to define annotation
operating points.

\textbf{Closed-form pacing.} We define the CSCS pacing rule as the
closed-form mapping from DCR and $\alpha_{\text{eff}}$ to the selection
parameter $\gamma$. This rule is designed around three requirements:
(i)~neutrality when no structural signal is available, (ii)~directional
adaptation when DCR is informative, and (iii)~gradual strengthening of
the adaptation as the effective budget increases:
\begin{equation}
    \gamma =
    0.5 + \frac{\text{DCR}}{4} \cdot
    \frac{\alpha_{\text{eff}}}{1 + \alpha_{\text{eff}}}
    \label{eq:gamma}
\end{equation}
DCR determines the direction of the trade-off between typicality and
uncertainty, while $m = \alpha_{\text{eff}}/(1+\alpha_{\text{eff}})
\in (0,1)$ modulates its strength. The divisor~4 keeps the CSCS pacing
rule in a bounded intermediate regime, with
$\gamma \in (0.25, 0.75)$ for all $\mathrm{DCR} \in [-1,1]$, avoiding
degenerate extremes while preserving directional adaptation. In practice,
the observed values remain within a conservative intermediate range.
When $\mathrm{DCR}=0$, Eq.~\ref{eq:gamma} reduces to $\gamma=0.5$
regardless of budget, and the score defaults to the unweighted geometric
mean. When $\mathrm{DCR}>0$, $\gamma$ increases above $0.5$, reflecting
that difficult samples tend to be representative; when
$\mathrm{DCR}<0$, $\gamma$ decreases below $0.5$, reflecting that
uncertain samples are more likely peripheral, and the score shifts toward
typicality-driven selection.

The CSCS pacing rule in Eq.~\ref{eq:gamma} is applied identically across
all datasets and budgets. Once the design constants ($\sqrt{N}$
normalization, divisor of~4, $k{=}20$, $s_{\min}{=}3$) are fixed, no
per-dataset hyperparameter tuning and no held-out calibration set are
required. This continuous adaptation of $\gamma$ distinguishes CSCS from
Hacohen's two-phase switch~\cite{hacohen2022budget}, HaCon's pseudo-label
selection rule~\cite{chen2023hacon}, and CSAL-3D's fixed hierarchical
scheme~\cite{zhu2025csal3d}, which rely on fixed or discretized
selection regimes rather than continuously adapting the trade-off
between typicality and uncertainty from pool-level statistics.

\begin{algorithm}[]
\caption{CSCS: Curriculum-Stratified Cold-Start Selection}
\label{alg:cscs}
\begin{algorithmic}[1]
\Require Unlabeled pool $\mathcal{U}=\{X_1,\dots,X_N\}$, budget $B$, SSL encoder $\Phi$
\Ensure Initial labeled set $\mathcal{L}_0$ with $|\mathcal{L}_0|=B$

\State Pretrain $\Phi$ on $\mathcal{U}$ and extract, for each $X_i$,
$\mathbf{z}_i$, $U(X_i)$, and $T(X_i)$
\State Normalize $U$ and $T$ to $[0,1]$
\State Compute $\mathrm{DCR} \gets \rho_s(\{U(X_i)\}, \{T(X_i)\})$
\State Cluster $\{\mathbf{z}_i\}$ into $B$ groups via $k$-means++
\State Merge clusters with size $< s_{\min}$ into their nearest neighbors
\State Compute $\alpha_{\text{eff}} \gets B/\sqrt{N}$
\State Compute $\gamma \gets
0.5 + \frac{\mathrm{DCR}}{4}\cdot
\frac{\alpha_{\text{eff}}}{1+\alpha_{\text{eff}}}$
\For{each resulting cluster $C_m$}
    \State Select $X_m^* \gets \argmax_{X_i \in C_m} T(X_i)^{1-\gamma} U(X_i)^\gamma$
\EndFor
\State $\mathcal{L}_0 \gets \{X_m^*\}_m$
\If{$|\mathcal{L}_0| < B$}
    \State Add remaining samples by global score ordering until $|\mathcal{L}_0|=B$
\EndIf
\State \Return $\mathcal{L}_0$
\end{algorithmic}
\end{algorithm}

\section{Experiments}
\label{sec:experiments}

\subsection{Datasets}

We evaluate CSCS on four 3D medical image segmentation benchmarks
spanning diverse pool sizes, modalities, and anatomical complexity.
From the Medical Segmentation Decathlon~\cite{antonelli2022msd}: (1)~\textbf{BraTS}, multi-parametric MRI (T1, T1ce, T2, FLAIR), 387/97 train/val split, evaluated using the standard composite tumor regions: whole tumor (WT), tumor core (TC), and enhancing tumor (ET); (2)~\textbf{Spleen}, CT, 34/8 split, single-organ segmentation. We also evaluate on
(3)~\textbf{FeTA}~\cite{payette2021fetal}, T2-weighted fetal brain MRI,
39/10 split, 7-class anatomical segmentation; and (4)~\textbf{DIANE},
an in-house placenta-fetal MRI dataset of 23 T2-weighted volumes
(18/5 split) acquired at 3T at CHRU-Nancy (NCT04328532) for segmentation of
the placenta and full fetal body (2~classes).

Annotation budgets are fixed fractions of each training pool
(20\%, 30\%, 40\%); the corresponding number of selected volumes $B$
is obtained by rounding to the nearest feasible integer
(Table~\ref{tab:datasets}). This evaluates all datasets at comparable
relative annotation levels while letting the absolute count vary with
pool size. We focus on these low-to-mid regimes because cold-start
initialization matters most when labeled data are scarce; at higher
budgets, differences between strategies narrow as coverage increases.
The effective budget $\alpha_{\text{eff}}=B/\sqrt{N}$ is reported
because it is used internally by the pacing rule, not as an additional
experimental budget.

\begin{table}[t]
\centering
\caption{Dataset statistics, annotation budgets (20\%/30\%/40\% of
the training pool), and DCR values computed from SSL embeddings.
$\gamma$ is computed by Eq.~\ref{eq:gamma} with no dataset-specific
tuning.}
\label{tab:datasets}
\scriptsize
\setlength{\tabcolsep}{3.5pt}
\begin{tabular}{lcccccccc}
\toprule
Dataset & $N$ & Mod. & Cls. & $B$ & $\alpha$ &
$\alpha_{\text{eff}}$ & DCR & $\gamma$ \\
\midrule
\multirow{3}{*}{BraTS}
  & \multirow{3}{*}{387} & \multirow{3}{*}{MRI} & \multirow{3}{*}{3}
  &  77 & 0.20 & 3.91 & \multirow{3}{*}{$-0.04$} & 0.49 \\
  & & & & 116 & 0.30 & 5.90 &  & 0.49 \\
  & & & & 155 & 0.40 & 7.88 &  & 0.49 \\
\midrule
\multirow{3}{*}{FeTA}
  & \multirow{3}{*}{39} & \multirow{3}{*}{MRI} & \multirow{3}{*}{7}
  &  8 & 0.21 & 1.28 & \multirow{3}{*}{$+0.23$} & 0.53 \\
  & & & & 12 & 0.31 & 1.92 &  & 0.54 \\
  & & & & 16 & 0.41 & 2.56 &  & 0.54 \\
\midrule
\multirow{3}{*}{Spleen}
  & \multirow{3}{*}{34} & \multirow{3}{*}{CT} & \multirow{3}{*}{1}
  &  7 & 0.21 & 1.20 & \multirow{3}{*}{$+0.29$} & 0.54 \\
  & & & & 10 & 0.29 & 1.71 &  & 0.55 \\
  & & & & 14 & 0.41 & 2.40 &  & 0.55 \\
\midrule
\multirow{3}{*}{DIANE}
  & \multirow{3}{*}{18} & \multirow{3}{*}{MRI} & \multirow{3}{*}{2}
  &  5 & 0.28 & 1.18 & \multirow{3}{*}{$+0.68$} & 0.59 \\
  & & & &  7 & 0.39 & 1.65 &  & 0.61 \\
  & & & &  9 & 0.50 & 2.12 &  & 0.62 \\
\bottomrule
\end{tabular}
\end{table}

\subsection{Implementation Details}
\label{subsec:implementation}

\textbf{Preprocessing.} CT intensities were clipped to
$[-1024, 1024]$~HU and normalized to $[0,1]$; MRI volumes underwent
z-score normalization after 1st--99th percentile clipping. No
dataset-specific preprocessing was applied beyond these standard steps.

\textbf{SSL pretraining.} We use the Swin-UNETR masked inpainting
framework of Tang et al.~\cite{tang2022swinunetr} trained on each
unlabeled pool with patch size $128^3$, masking ratio~0.3, and AdamW
(lr~$10^{-4}$, 5\,000 iterations). Volume-level embeddings
($d{=}768$) are obtained via global average pooling and L2
normalization, and $U(X_i)$ and $T(X_i)$ ($k{=}20$) are extracted from
this frozen encoder (Section~\ref{sec:ssl}). The same encoder,
embeddings, normalization, and KNN definition are shared across all
methods, isolating the contribution of the selection strategy.

\textbf{Segmentation training.} Selected sets train nnU-Net in its 3D
full-resolution configuration~\cite{isensee2021nnunet} with the
Ranger21 optimizer~\cite{wright2021ranger21} (lr~$10^{-3}$, 250
epochs). Each result is averaged over 3 independent runs on a fixed
split, each capturing both selection randomness (k-means
initialization, tie-breaking) and downstream training randomness,
following recommendations for robust low-data
evaluation~\cite{maier2024metrics}. The fully supervised nnU-Net is the
upper bound. The CSCS pacing rule uses the same design constants
throughout ($\sqrt{N}$ normalization, divisor~4, $k{=}20$,
$s_{\min}{=}3$), with no per-dataset tuning or held-out calibration.
All experiments ran on a single NVIDIA A100 GPU with
PyTorch~2.3.1~\cite{paszke2019pytorch} and MONAI~1.3.1~\cite{cardoso2022monai}.

\subsection{Compared Methods}

We compare CSCS against five baselines covering representative,
diversity-based, and uncertainty-aware cold-start behaviors; the goal
is a comparison against representative fixed strategies rather than an
exhaustive sweep of heuristics. For cluster-based methods the same
k-means partition is reused to isolate scoring effects:
(1)~\textbf{Random}, uniform sampling;
(2)~\textbf{TypiClust}~\cite{hacohen2022budget}, most-typical sample
per cluster (representativeness-dominant end of the continuum);
(3)~\textbf{FPS}~\cite{jin2022oneshot}, farthest point sampling for
maximum pairwise diversity;
(4)~\textbf{ProbCover}~\cite{yehuda2022probcover}, greedy ball-covering
maximizing covered probability mass;
(5)~\textbf{CSAL-3D}~\cite{zhu2025csal3d}, an uncertainty-first
cluster-based baseline using a fixed hierarchical procedure.
Methods that jointly optimize representation and selection
(e.g.\ VAAL~\cite{sinha2019vaal}, USL~\cite{wang2022usl}) are excluded
as they prevent fair isolation of the selection contribution. An
additional control (\textbf{Random-per-cluster}) is used in the
ablation (Section~\ref{sec:ablation}) to separate diversity from
informed scoring.

\subsection{Evaluation Metrics}
\label{subsec:metrics}

To evaluate the performance of the proposed cold-start selection strategy for 3D medical image segmentation, we used complementary metrics assessing both regional overlap and boundary accuracy. For BraTS, metrics were computed on the standard composite tumor regions WT, TC, and ET, following common BraTS evaluation practice. For FeTA and DIANE, metrics were computed independently for each foreground class and then averaged across classes; for Spleen, the single foreground class was used. Individual BraTS label-wise results for NCR/NET, edema, and ET are reported only in the Supplementary Material.

The Dice similarity coefficient (DSC) was used as the primary
region-overlap metric. It measures the similarity between the predicted
segmentation and the corresponding ground-truth annotation. For a
predicted region $P$ and a ground-truth region $G$, the DSC is defined as
\begin{equation}
    \mathrm{DSC}(P,G) =
    \frac{2|P \cap G|}{|P| + |G|},
    \label{eq:dice}
\end{equation}
where $|P \cap G|$ denotes the number of correctly segmented voxels,
and $|P|$ and $|G|$ denote the number of voxels in the predicted and
ground-truth regions, respectively. DSC ranges from 0 to 1, with higher
values indicating stronger spatial overlap. In the main results, DSC is reported as a percentage and averaged over the relevant evaluation regions: WT/TC/ET for BraTS, all foreground classes for FeTA and DIANE, and the single foreground class for Spleen.

Boundary accuracy was assessed using the 95th-percentile Hausdorff
distance (HD95). Let $\partial P$ and $\partial G$ denote the sets of
surface points of the predicted and ground-truth regions, respectively,
and let $d(p,g)$ be the Euclidean distance between two surface points
$p$ and $g$. The directed distance from a point $p \in \partial P$ to
the surface $\partial G$ is defined as
\begin{equation}
    d(p,\partial G) = \min_{g \in \partial G} d(p,g).
\end{equation}
The HD95 is then computed as the maximum of the 95th percentiles of the
bidirectional surface distances:
\begin{equation}
    \mathrm{HD95}(P,G) =
    \max \left\{
    \operatorname{percentile}_{95}
    \left( \{ d(p,\partial G) \;|\; p \in \partial P \} \right),
    \operatorname{percentile}_{95}
    \left( \{ d(g,\partial P) \;|\; g \in \partial G \} \right)
    \right\}.
    \label{eq:hd95}
\end{equation}
Unlike the maximum Hausdorff distance, which is highly sensitive to
isolated outlier points, HD95 reduces the influence of spurious distant
voxels and provides a more robust estimate of boundary mismatch.
Lower HD95 values indicate more accurate boundary delineation.

In addition to DSC and HD95, the Intersection-over-Union (IoU) was
reported as a complementary overlap metric:
\begin{equation}
\mathrm{IoU}(P,G) =
\frac{|P \cap G|}{|P \cup G|}.
\label{eq:iou}
\end{equation}
IoU ranges from 0 to 1, with higher values indicating better agreement
between prediction and ground truth. For a given evaluation region or
class, IoU is monotonically related to DSC:
\begin{equation}
\mathrm{IoU} = \frac{\mathrm{DSC}}{2-\mathrm{DSC}},
\end{equation}
where DSC is expressed as a fraction between 0 and 1. For multi-region
or multi-class datasets, IoU was computed independently for each
evaluation region or class and then averaged arithmetically, following
the same aggregation convention as Dice and HD95. Thus, the reported
mean IoU corresponds to the mean of region-wise or class-wise IoU
values, and not to the Dice-to-IoU transform applied to an already
averaged Dice score. IoU is therefore interpreted as a complementary
overlap metric rather than as independent evidence.

Finally, per-class Precision and Recall are reported in the
Supplementary material to further characterize the error profile of each
method. Precision measures the proportion of predicted foreground voxels
that are correct, whereas Recall measures the proportion of ground-truth
foreground voxels recovered by the model:
\begin{equation}
    \mathrm{Precision}(P,G) =
    \frac{|P \cap G|}{|P|},
    \qquad
    \mathrm{Recall}(P,G) =
    \frac{|P \cap G|}{|G|}.
    \label{eq:precision_recall}
\end{equation}
These metrics decompose overlap errors into false-positive and
false-negative contributions and are therefore useful for identifying
whether a method tends to over-segment or under-segment specific
structures.

\subsection{Quantitative and Qualitative Analysis}
\label{sec:main_results}

\begin{figure}[!t]
\centering
\includegraphics[width=\linewidth]{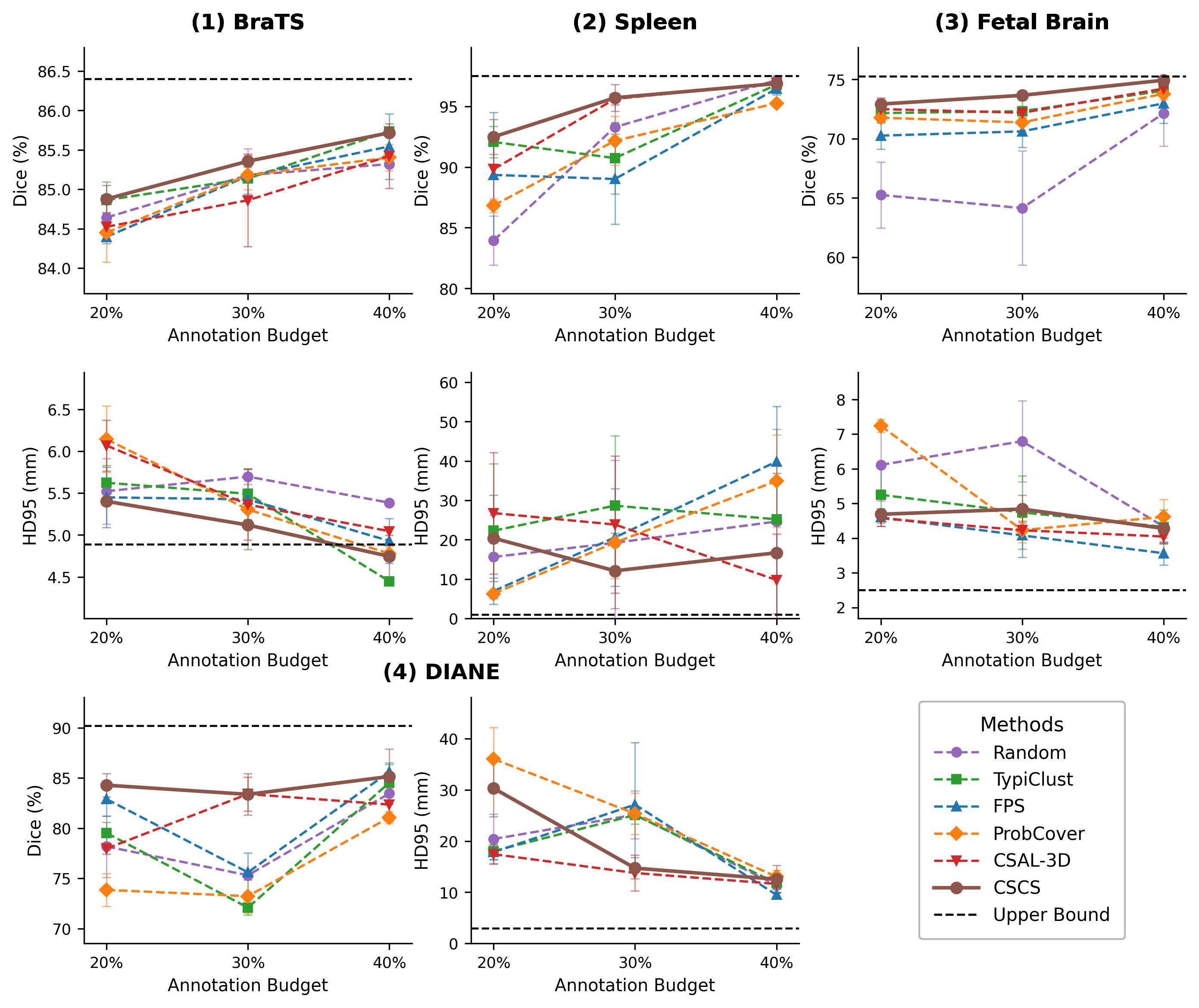}
\caption{Segmentation performance as a function of annotation budget
across four benchmarks. \textbf{Top rows}: mean Dice~(\%) and HD95~(mm)
for BraTS, Spleen, and FeTA. \textbf{Bottom row}: DIANE dataset.
Error bars: $\pm$1 standard deviation over 3 independent runs.}
\label{fig:curves}
\end{figure}

Table~\ref{tab:main_results} reports mean Dice~(\%) at the three
budgets, with absolute counts $B$ in parentheses;
Fig.~\ref{fig:curves} shows the Dice and HD95 trends. Additional detailed results, including IoU, HD95, and per-run standard
deviations for all datasets and budgets, are provided in the
Supplementary material. Given the
limited number of runs, we emphasize consistent trends and a
distribution-free statistical analysis rather than isolated per-cell
comparisons.

\begin{table}[t]
\centering
\caption{Mean Dice~(\%) over 3 independent runs. For BraTS: composite
Dice (mean of WT, TC, ET regions); for other datasets: mean over
foreground classes. Budgets correspond to 20\%/30\%/40\% of each
training pool, with absolute counts $B$ shown in parentheses.
Best in \textbf{bold}, second-best \underline{underlined}.}
\label{tab:main_results}
\scriptsize
\setlength{\tabcolsep}{2.5pt}
\begin{tabular}{l ccc ccc ccc ccc}
\toprule
& \multicolumn{3}{c}{\textbf{BraTS} ($N{=}387$)}
& \multicolumn{3}{c}{\textbf{FeTA} ($N{=}39$)}
& \multicolumn{3}{c}{\textbf{Spleen} ($N{=}34$)}
& \multicolumn{3}{c}{\textbf{DIANE} ($N{=}18$)} \\
\cmidrule(lr){2-4} \cmidrule(lr){5-7}
\cmidrule(lr){8-10} \cmidrule(lr){11-13}
& \scriptsize 20\% & \scriptsize 30\% & \scriptsize 40\%
& \scriptsize 20\% & \scriptsize 30\% & \scriptsize 40\%
& \scriptsize 20\% & \scriptsize 30\% & \scriptsize 40\%
& \scriptsize 20\% & \scriptsize 30\% & \scriptsize 40\% \\
& \scriptsize (77) & \scriptsize (116) & \scriptsize (155)
& \scriptsize (8)  & \scriptsize (12)  & \scriptsize (16)
& \scriptsize (7)  & \scriptsize (10)  & \scriptsize (14)
& \scriptsize (5)  & \scriptsize (7)   & \scriptsize (9) \\
\midrule
Random
& 84.6 & 85.2 & 85.3
& 65.2 & 64.1 & 72.1
& 84.0 & 93.3 & \textbf{97.2}
& 78.2 & 75.3 & 83.5 \\

TypiClust
& \underline{84.9} & 85.1 & \textbf{85.7}
& 72.2 & \underline{72.3} & 74.0
& \underline{92.1} & 90.7 & 96.8
& 79.5 & 72.1 & 84.5 \\

FPS
& 84.4 & 85.2 & 85.5
& 70.3 & 70.6 & 73.0
& 89.4 & 89.0 & 96.4
& \underline{82.9} & 75.6 & \textbf{85.6} \\

ProbCover
& 84.5 & \underline{85.2} & 85.4
& 71.8 & 71.4 & 73.8
& 86.8 & 92.2 & 95.2
& 73.9 & 73.2 & 81.1 \\

CSAL-3D
& 84.5 & 84.9 & 85.4
& \underline{72.5} & 72.2 & \underline{74.2}
& 89.8 & \underline{95.6} & \underline{96.9}
& 78.0 & \textbf{83.4} & 82.4 \\

\midrule
\rowcolor{gray!10}
\textbf{CSCS}
& \textbf{84.9} & \textbf{85.4} & \underline{85.7}
& \textbf{72.9} & \textbf{73.7} & \textbf{74.9}
& \textbf{92.5} & \textbf{95.7} & \underline{96.9}
& \textbf{84.3} & \underline{83.4} & \underline{85.1} \\
\midrule
\textit{Full Sup.}
& 86.4 & \multicolumn{2}{c}{---}
& 75.3 & \multicolumn{2}{c}{---}
& 97.5 & \multicolumn{2}{c}{---}
& 90.2 & \multicolumn{2}{c}{---} \\
\bottomrule
\end{tabular}
\end{table}

\begin{figure}[!t]
\centering
\includegraphics[width=\linewidth]{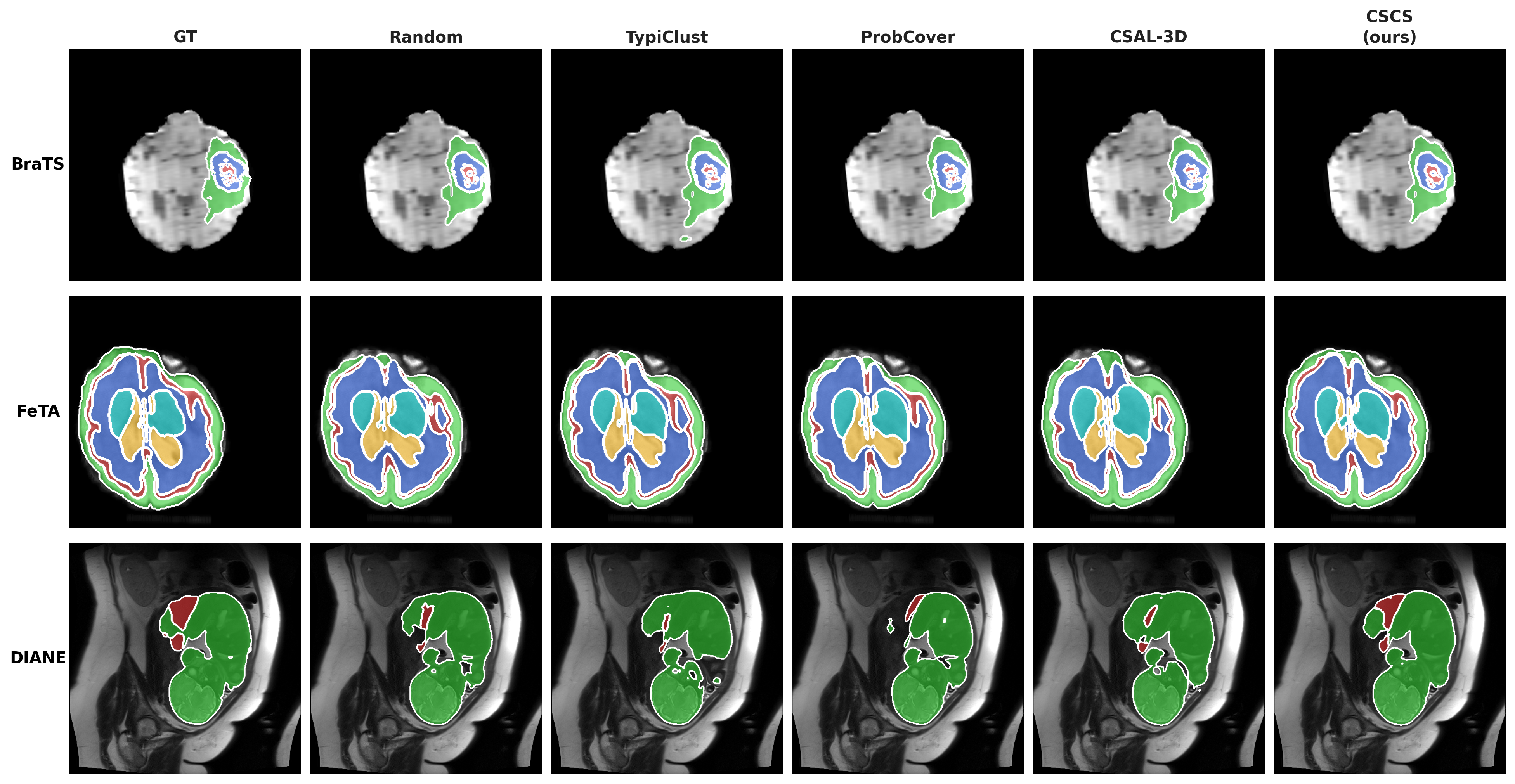}
\caption{Qualitative segmentation examples on the BraTS, FeTA and
DIANE datasets at 20\% annotation budget ($B{=}77$, $B{=}8$, and
$B{=}5$, respectively).}
\label{fig:qualitative}
\end{figure}

\textbf{Reducing expert annotation effort.}
Because each 3D volume requires tens of minutes of specialist
delineation~\cite{tajbakhsh2020embracing,wang2021annotation}, the
practical value of cold-start selection is the quality obtained per
annotated volume. CSCS approaches the fully supervised upper bound
while annotating only a fraction of the pool: 73.7\% Dice at 31\% budget on FeTA (vs.\ 75.3\% full) and 85.4\%
composite Dice at 30\% on BraTS (vs.\ 86.4\%), i.e.\ within
$1$--$1.6$~pp of the bound while leaving
most volumes unannotated. The benefit is largest at the lowest budgets
and smallest pools, where each avoided annotation carries the highest
marginal cost; on DIANE ($N{=}18$), a strong initial set of $B{=}5$
volumes already yields 84.3\% Dice. By choosing this set well in a
single pre-annotation step, CSCS concentrates expert effort on the
volumes that most improve the first model.

\textbf{Budget regimes.}
At the lowest budget (20\%), where selection matters most, CSCS attains
the highest or tied-highest Dice on all four datasets; the clearest
margin is on DIANE (84.3\% vs.\ 82.9\% for FPS, $+1.4$~pp), with FeTA
72.9\% (vs.\ 72.5\% for CSAL-3D) and BraTS showing a near-tie between
CSCS and TypiClust after rounding (84.9\% for both methods).
At 30\%, CSCS remains among the top two methods on all four datasets,
with the clearest gain on FeTA (73.7\% vs.\ 72.3\%) and near-tied
performance with CSAL-3D on DIANE. Fixed strategies are more
dataset-specific here, e.g.\ TypiClust is strong on BraTS but drops on
DIANE, whereas CSAL-3D is strong on Spleen/DIANE but weaker on FeTA.
At 40\%, differences narrow as coverage grows: CSCS remains first on
FeTA and close to the best method on BraTS and DIANE, while all methods
approach the upper bound on Spleen. These shifts match the adaptive
rule: near-zero DCR (BraTS) keeps selection balanced, whereas higher
positive DCR (DIANE) tilts it toward difficulty-aware selection.

\textbf{Cross-dataset consistency and statistical analysis.}
To summarize behavior across the 12 dataset--budget configurations
without over-interpreting small per-cell differences, we rank the six
methods within each configuration (rank~1 = best Dice; ties broken by
mean Dice) and aggregate the ranks. The full per-configuration rank matrix is provided in the Supplementary material. CSCS attains the
best mean rank across the 12 configurations, is first in 8 and
first-or-second in 11, and never falls below rank~3, whereas every
fixed baseline drops to rank~6 at least once. A Friedman test rejects
equal performance ($\chi^2_F\approx23.5$, df${=}5$, $p<0.001$), and the
Nemenyi post-hoc ($\mathrm{CD}\approx2.18$) finds CSCS significantly
better than Random, FPS, and ProbCover, while statistically comparable
to TypiClust and CSAL-3D. As the three budgets of a dataset share the
same pool, these tests are indicative rather than definitive, but they
provide a more principled summary than per-cell comparisons: CSCS
matches the strongest fixed baselines while being significantly more
reliable than the weaker ones, using a single rule rather than one that
happens to suit a given dataset.

The boundary metric is consistent with this picture: HD95
(Fig.~\ref{fig:curves}) is lowest or near-lowest for CSCS across
datasets, and the qualitative examples (Fig.~\ref{fig:qualitative})
show improved delineation of small structures and ambiguous boundaries.
The non-monotonic behavior on DIANE is expected given the very small
pool, where nearby budgets correspond to discrete changes in the
selected volumes rather than smooth increases in coverage; the relevant
criterion there is the absence of large drops, which CSCS satisfies.

\textbf{Summary.}
Gains are not uniformly large in every configuration, but they are most
consistent in the low-to-mid budget settings where initialization
matters most, particularly on FeTA and DIANE, and narrower on BraTS and
high-budget Spleen. Supported by the rank analysis, this matches our
central claim: the value of CSCS lies less in large per-case gains than
in avoiding large drops across heterogeneous datasets and budgets.

\subsection{Ablation Study}
\label{sec:ablation}

We evaluate core design choices through targeted ablations, all run at
30\% budget over 3 independent runs.

\textbf{Effect of $\gamma$ and the composite score.}
Table~\ref{tab:ablation_gamma_dcr} compares the predicted $\gamma$
(Eq.~\ref{eq:gamma}) against fixed values
$\gamma \in \{0.0, 0.3, 0.5, 0.7, 1.0\}$. Two points emerge. First, the extremes are not robust and can fail
badly, e.g.\ on Spleen pure uncertainty ($\gamma{=}1$) yields 90.8\%
Dice / 83.5~mm HD95 against 95.7\% / 12.1~mm for CSCS; pure
typicality ($\gamma{=}0$, equivalent to TypiClust) also underperforms
on three of four datasets. Second, no single fixed intermediate
$\gamma$ dominates: $\gamma{=}0.5$ leads on BraTS and Spleen,
$\gamma{=}0.7$ on DIANE, $\gamma{=}0.3$ on FeTA. The predicted $\gamma$ stays within $0.2$~pp of the best fixed value on
BraTS ($-0.2$~pp vs.\ $\gamma{=}0.5$, 85.4\% vs.\ 85.6\%) and remains
competitive on Spleen ($-1.1$~pp), while it achieves the best Dice and
HD95 on DIANE. Overall, the DCR-driven calibration finds a robust
operating point without knowing the best fixed $\gamma$ in advance.

\begin{table}[t]
\centering
\caption{Ablation: effect of $\gamma$ at 30\% budget.
Mean Dice~(\%) and HD95~(mm) over 3 independent runs
(2 for $\gamma{=}1.0$, BraTS). For BraTS: composite Dice
(mean of WT, TC, ET); $\gamma{=}0.0$ (T-only) is equivalent
to TypiClust by construction, results taken from main experiments.
Best in \textbf{bold}, second-best \underline{underlined}.}
\label{tab:ablation_gamma_dcr}
\scriptsize
\setlength{\tabcolsep}{3pt}
\begin{tabular}{l cc cc cc cc}
\toprule
& \multicolumn{2}{c}{\textbf{BraTS}}
& \multicolumn{2}{c}{\textbf{FeTA}}
& \multicolumn{2}{c}{\textbf{Spleen}}
& \multicolumn{2}{c}{\textbf{DIANE}} \\
\cmidrule(lr){2-3}\cmidrule(lr){4-5}
\cmidrule(lr){6-7}\cmidrule(lr){8-9}
$\gamma$ strategy
& Dice & HD95 & Dice & HD95 & Dice & HD95 & Dice & HD95 \\
\midrule
$\gamma{=}0.0$ (T-only)
  & 85.1 & 5.49 & 71.8 & 4.2 & 88.9 & 38.1 & 72.1 & 25.1 \\
$\gamma{=}0.3$
  & 85.3 & \underline{5.06} & \textbf{74.2} & \textbf{3.5}
  & 91.9 & \underline{9.0} & 79.8 & 31.7 \\
$\gamma{=}0.5$
  & \textbf{85.6} & \textbf{5.03} & \underline{73.9} & 4.5
  & \textbf{96.8} & \textbf{1.6} & 82.2 & 19.7 \\
$\gamma{=}0.7$
  & 85.2 & \underline{5.06} & 73.9 & \underline{4.3}
  & 92.3 & 61.9 & \underline{82.8} & \underline{16.7} \\
$\gamma{=}1.0$ (U-only)
  & 85.3 & 5.26 & 67.4 & 5.6
  & 90.8 & 83.5 & 77.8 & 23.2 \\
\midrule
\rowcolor{gray!10}
Predicted $\gamma$ (Eq.~\ref{eq:gamma})
  & \underline{85.4} & 5.12
  & 73.7 & 4.6
  & \underline{95.7} & 12.1
  & \textbf{83.4} & \textbf{14.7} \\
\bottomrule
\end{tabular}
\end{table}

\textbf{Effect of DCR.}
Setting DCR to zero reduces Eq.~\ref{eq:gamma} to $\gamma{=}0.5$
(the $\gamma{=}0.5$ row). On BraTS (DCR${=}{-}0.04$) this is negligible
since the prediction is already near 0.5; on DIANE (DCR${=}{+}0.68$)
the DCR-aware prediction yields 83.4\% vs.\ 82.2\% for $\gamma{=}0.5$
($+1.2$~pp Dice, $-5.0$~mm HD95). Curriculum adaptation is thus driven
primarily by dataset geometry via DCR, with the budget factor acting as
a conservative modulator. While DCR is a coarse global descriptor,
these results suggest it captures a useful aspect of the alignment
between uncertainty and representativeness for guiding cold-start
selection.

\textbf{Diversity control: Random-per-cluster.}
Table~\ref{tab:ablation_diversity} isolates informed scoring from the
diversity guarantee of clustering. Random-per-cluster selects one
sample per cluster with no scoring. CSCS substantially outperforms it on three of four datasets
($+8.5$~pp DIANE, $+3.6$~pp Spleen, $+0.4$~pp BraTS); on FeTA the
two are nearly identical (73.7\% vs.\ 73.6\%), so there diversity
explains most of the gain. Notably, Random-per-cluster sometimes
underperforms unconstrained Random (DIANE 74.9\% vs.\ 75.3\%;
BraTS 85.0\% vs.\ 85.2\%), showing that enforcing
diversity without informed scoring can isolate low-quality regions;
CSCS avoids this by combining structural diversity with the composite
score.

\begin{table}[t]
\centering
\caption{Ablation: diversity contribution at 30\% budget.
Mean Dice~(\%) and HD95~(mm) over 3 independent runs.
For BraTS: composite Dice (mean of WT, TC, ET).
Best in \textbf{bold}.}
\label{tab:ablation_diversity}
\scriptsize
\setlength{\tabcolsep}{3pt}
\begin{tabular}{l cc cc cc cc}
\toprule
& \multicolumn{2}{c}{\textbf{BraTS}}
& \multicolumn{2}{c}{\textbf{FeTA}}
& \multicolumn{2}{c}{\textbf{Spleen}}
& \multicolumn{2}{c}{\textbf{DIANE}} \\
\cmidrule(lr){2-3}\cmidrule(lr){4-5}
\cmidrule(lr){6-7}\cmidrule(lr){8-9}
Configuration
& Dice & HD95 & Dice & HD95 & Dice & HD95 & Dice & HD95 \\
\midrule
Random (no structure)
  & 85.2 & 5.70 & 64.1 & 6.8 & 93.3 & 19.2 & 75.3 & 25.1 \\
Random-per-cluster
  & 85.0 & 5.55 & 73.6 & \textbf{4.4}
  & 92.1 & 45.0 & 74.9 & 27.4 \\
\midrule
\rowcolor{gray!10}
CSCS (ours)
  & \textbf{85.4} & \textbf{5.12}
  & \textbf{73.7} & 4.8
  & \textbf{95.7} & \textbf{12.1}
  & \textbf{83.4} & \textbf{14.7} \\
\bottomrule
\end{tabular}
\end{table}

\textbf{Summary.}
These ablations support three conclusions: (i)~fixed extreme regimes
are not robust for cold-start selection; (ii)~the best balance is
dataset-dependent, and DCR provides useful pool-level information for
adjusting it; and (iii)~diversity alone does not explain the gains of
CSCS---informed scoring within clusters is essential on most
benchmarks.

\subsection{Discussion}
\label{sec:discussion}

Our results point to a simple conclusion: in the absence of labels,
acquisition should be matched to the geometry of the unlabeled pool
rather than driven by a fixed criterion. This is consistent with budget-aware active
learning theory, which shows that opposite querying regimes can be
preferable at different budgets and that the best strategy depends on
both the problem and the annotation
regime~\cite{hacohen2022budget,hacohen2023select}. CSCS should
therefore be interpreted as a practical instantiation of
dataset-adaptive cold-start selection rather than a theoretically
optimal rule.

This interpretation clarifies the role of DCR. We do not view DCR as a
unique or exhaustive descriptor of pool structure, but as a coarse
global proxy for whether difficulty aligns with representativeness in
the unlabeled pool. When this alignment is weak, as on BraTS
(DCR${=}{-}0.04$), $\gamma$ remains near $0.5$ and selection is
effectively balanced; when it is stronger, as on DIANE
(DCR${=}{+}0.68$), the rule shifts toward difficulty-aware selection,
and the ablation confirms a corresponding +1.2~pp gain over the
balanced baseline. DCR does not fully explain dataset behavior. Its value is more
modest: it provides useful pre-annotation signal for avoiding
poorly suited fixed strategies.

It is also important to distinguish the two levels at which CSCS
operates. DCR calibrates the global trade-off parameter $\gamma$; it
does not rank or select samples directly. The actual selection decision
remains cluster-wise through maximization of the composite score~$S$.
A coarse global statistic suffices here: it sets the balance between
two signals, while the actual selection is handled locally per cluster. As reported in Table~\ref{tab:datasets}, $\gamma$ stays
within a moderate range ($0.49$--$0.62$) across all experiments,
confirming that the rule operates in a conservative intermediate regime
rather than swinging between extremes.

The main empirical message is one of robustness rather than uniformly
large gains. Improvements are modest in near-saturation settings, where
most methods approach the fully supervised upper bound, but CSCS more
consistently avoids the failure modes of fixed strategies: overly
conservative behavior when difficult samples are informative, and
overly aggressive behavior when uncertainty is dominated by peripheral
cases. In this setting, avoiding large performance drops across
heterogeneous regimes is a more meaningful success criterion than
isolated wins in a few configurations.

Our findings also reinforce the importance of the underlying
representation. Both typicality and reconstruction-based uncertainty
are derived from SSL features, so their usefulness depends on how well
the unlabeled pool is organized in the embedding space. CSCS should
therefore be understood as a representation-dependent cold-start
strategy, and investigating the sensitivity of the pacing rule to
different SSL backbones is a natural direction for future work.

Several limitations are worth noting. The pacing rule rests on
fixed design constants ($\sqrt{N}$ normalization, divisor of~4),
whose functional form has a natural geometric interpretation:
$\beta{=}1/2$ is the midpoint of the family $B/N^\beta$
(Eq.~\ref{eq:alpha_family}), and the divisor~4 keeps $\gamma$
within $(0.25, 0.75)$. These constants are not theoretically
optimal and
other choices may prove effective in different settings.
Additionally, the current validation covers four datasets and three
independent runs, which limits the statistical strength of the
conclusions; broader evaluation across more diverse pools and
downstream frameworks would strengthen the generality of the
findings. Finally, our evaluation focuses on one-shot
initialization; extending dataset-aware pacing to iterative active
learning remains an open and promising direction, despite recent
progress~\cite{luth2025nnactive,traub2026clasp}. More broadly, the
present results support a principle --- adapting acquisition to pool
geometry before annotation --- rather than advocating for any single
universally preferred rule.

\section{Conclusion}

We addressed the cold-start problem in active learning for 3D medical
image segmentation by introducing CSCS, a dataset-adaptive one-shot
initialization strategy that balances representativeness and difficulty
using pool-level statistics available before annotation. CSCS does not advocate a fixed query criterion. It operationalizes
the idea that cold-start acquisition should depend on both annotation
budget and pool geometry.

Across four heterogeneous benchmarks, CSCS ranks among the
top-performing methods across all evaluated configurations, with stronger benefits in low-to-mid annotation regimes where initialization
matters most. More broadly, our results suggest that effective
cold-start selection benefits from adapting the trade-off between
typicality and uncertainty to the structure of the unlabeled pool,
rather than relying on a single fixed acquisition rule. Future work
will explore richer pool descriptors beyond the global DCR, extensions
to iterative active learning, and sensitivity of the pacing rule to
different self-supervised backbones.

\section*{CRediT authorship contribution statement}

\textbf{Rémi Hattat:} Conceptualization, Methodology, Software, Validation, Formal analysis, Investigation, Data curation,  Writing -- original draft, Writing -- review \& editing. \textbf{Marine Beaumont:} Investigation, Project administration, Writing -- review \& editing. \textbf{Charline Bertholdt:} Resources, Project administration, Writing -- review \& editing. \textbf{Gabriela Hossu:} Resources, Writing -- review \& editing. \textbf{Olivier Morel:} Supervision, Resources, Writing -- review \& editing. \textbf{Bailiang Chen:} Supervision, Resources, Conceptualization, Writing -- review \& editing.

\section*{Declaration of competing interest}
The authors declare that they have no known competing financial
interests or personal relationships that could have appeared to influence
the work reported in this paper.

\section*{Acknowledgments}
This work was conducted on a platform co-funded by the French government through the Contrat de Plan Etat-Région (CPER2015-2020 IT2MP) and by the
European Regional Development Fund (ERDF 2014-2020). The platform is affiliated
with the France Life Imaging (ANR-11-INBS-0006).

\section*{Supplementary Material}
Supplementary material is provided as an ancillary file accompanying this arXiv submission.

\section*{Data availability}
The source code for the proposed method is openly available in a
public repository at the following URL: \url{https://github.com/rhattat/CSCS-AL}. The public datasets used in this study are available from their
respective sources. The in-house DIANE imaging data are not publicly
available due to institutional and ethical restrictions.

\section*{Declaration of generative AI and AI-assisted technologies in the writing process}
During the preparation of this work, the authors used ChatGPT to assist
with language editing, manuscript structuring, and clarity improvement.
After using this tool, the authors reviewed and edited the content as
needed and take full responsibility for the content of the publication.

\bibliographystyle{splncs04}

\begin{thebibliography}{99}


\bibitem{isensee2021nnunet}
Isensee, F., Jaeger, P.F., Kohl, S.A., Petersen, J., Maier-Hein, K.H.:
nnU-Net: a self-configuring method for deep learning-based biomedical image segmentation.
Nature Methods \textbf{18}(2), 203--211 (2021)

\bibitem{hatamizadeh2022swinunetr}
Hatamizadeh, A., Nath, V., Tang, Y., Yang, D., Roth, H.R., Xu, D.:
Swin UNETR: Swin transformers for semantic segmentation of brain tumors in MRI images.
In: BrainLes@MICCAI. pp. 272--284 (2022)

\bibitem{tajbakhsh2020embracing}
Tajbakhsh, N., Jeyaseelan, L., Li, Q., Chiang, J.N., Wu, Z., Ding, X.:
Embracing imperfect datasets: A review of deep learning solutions for medical image segmentation.
Medical Image Analysis \textbf{63}, 101693 (2020)

\bibitem{wang2021annotation}
Wang, S., Li, C., Wang, R., Liu, Z., et al.:
Annotation-efficient deep learning for automatic medical image segmentation.
Nature Communications \textbf{12}(1), 5915 (2021)

\bibitem{wang2024survey}
Wang, H., Jin, Q., Li, S., Liu, S., Wang, M., Song, Z.:
A comprehensive survey on deep active learning in medical image analysis.
Medical Image Analysis \textbf{95}, 103201 (2024)


\bibitem{settles2009active}
Settles, B.:
Active learning literature survey.
Computer Sciences Technical Report 1648, University of Wisconsin-Madison (2009)

\bibitem{sener2018active}
Sener, O., Savarese, S.:
Active learning for convolutional neural networks: A core-set approach.
In: ICLR (2018)

\bibitem{gal2017deep}
Gal, Y., Islam, R., Ghahramani, Z.:
Deep Bayesian active learning with image data.
In: ICML. pp. 1183--1192 (2017)

\bibitem{ash2020badge}
Ash, J.T., Zhang, C., Krishnamurthy, A., Langford, J., Agarwal, A.:
BADGE: Batch active learning by diverse gradient embeddings.
In: ICLR (2020)

\bibitem{yehuda2022probcover}
Yehuda, O., Dekel, A., Hacohen, G., Weinshall, D.:
Active learning through a covering lens.
In: NeurIPS (2022)

\bibitem{huang2010querying}
Huang, S.J., Jin, R., Zhou, Z.H.:
Active learning by querying informative and representative examples.
In: NeurIPS. pp. 892--900 (2010)

\bibitem{sinha2019vaal}
Sinha, S., Ebrahimi, S., Darrell, T.:
Variational adversarial active learning.
In: ICCV. pp. 5972--5981 (2019)

\bibitem{wang2022usl}
Wang, X., Lian, L., Yu, S.X.:
Unsupervised Selective Labeling for More Effective Semi-Supervised Learning.
In: Avidan, S., Brostow, G., Ciss{\'e}, M., Farinella, G.M., Hassner, T. (eds.)
Computer Vision -- ECCV 2022. LNCS, vol. 13690, pp. 427--445.
Springer, Cham (2022)


\bibitem{liu2023colossal}
Liu, H., Li, H., Yao, X., Fan, Y., et al.:
COLosSAL: A benchmark for cold-start active learning for 3D medical image segmentation.
In: MICCAI. pp. 25--34 (2023)

\bibitem{chen2023hacon}
Chen, L., Bai, Y., Huang, S., Lu, Y., Wen, B., Yuille, A.L., Zhou, Z.:
Making Your First Choice: To Address Cold Start Problem in Medical Active Learning.
In: Medical Imaging with Deep Learning. Proceedings of Machine Learning Research,
vol. 227, pp. 496--525 (2024)

\bibitem{zhu2025csal3d}
Zhu, N., Ye, P., Zhong, L., Yue, Q., Zhang, S., Wang, G.:
CSAL-3D: Cold-start active learning for 3D medical image segmentation via SSL-driven uncertainty-reinforced diversity sampling.
In: MICCAI (2025)

\bibitem{hacohen2022budget}
Hacohen, G., Dekel, A., Weinshall, D.:
Active learning on a budget: Opposite strategies suit high and low budgets.
In: ICML. pp. 8175--8195 (2022)

\bibitem{hacohen2023select}
Hacohen, G., Weinshall, D.:
How to select which active learning strategy is best suited for your specific problem and budget.
In: NeurIPS. pp. 13395--13407 (2023)

\bibitem{ma2024acal}
Ma, S., Du, H., Curran, K.M., Lawlor, A., Dong, R.:
Adaptive Curriculum Query Strategy for Active Learning in Medical Image Classification.
In: Medical Image Computing and Computer Assisted Intervention -- MICCAI 2024.
pp. 48--57 (2024)

\bibitem{chandra2021initial}
Chandra, A.L., Desai, S.V., Devaguptapu, C., Balasubramanian, V.N.:
On Initial Pools for Deep Active Learning.
In: NeurIPS 2020 Workshop on Pre-registration in Machine Learning.
Proceedings of Machine Learning Research, vol. 148, pp. 14--32 (2021)

\bibitem{nath2022warm}
Nath, V., Yang, D., Roth, H.R., Xu, D.:
Warm-Start Active Learning with Proxy Labels and Selection via Semi-Supervised Fine-Tuning.
In: Medical Image Computing and Computer Assisted Intervention (MICCAI).
pp. 297--308 (2022)

\bibitem{yuan2024foundation}
Yuan, J., et al.:
Foundation Model Makes Clustering a Better Initialization for Cold-Start Active Learning.
arXiv preprint arXiv:2402.02561 (2024)

\bibitem{luth2025nnactive}
L\"uth, C.T., Traub, J., Kahl, K.-C., Bungert, T.J., Klein, L., Kr\"amer, L.,
Jaeger, P.F., Isensee, F., Maier-Hein, K.:
nnActive: A Framework for Evaluation of Active Learning in 3D Biomedical Segmentation.
Transactions on Machine Learning Research (2025)

\bibitem{traub2026clasp}
L\"uth, C.T., Traub, J., Kahl, K.-C., Bungert, T.J., Klein, L., Kr\"amer, L.,
Jaeger, P.F., Maier-Hein, K., Isensee, F.:
Finally Outshining the Random Baseline: A Simple and Effective Solution for Active Learning in 3D Biomedical Imaging.
Transactions on Machine Learning Research (2026)



\bibitem{antonelli2022msd}
Antonelli, M., Reinke, A., Bakas, S., et al.:
The Medical Segmentation Decathlon.
Nature Communications \textbf{13}(1), 4128 (2022)

\bibitem{payette2021fetal}
Payette, K., de Dumast, P., Kebiri, H., Ezhov, I., Paetzold, J.C., Shit, S., et al.:
An automatic multi-tissue human fetal brain segmentation benchmark using the fetal tissue annotation dataset.
Scientific Data \textbf{8}(1), 167 (2021)

\bibitem{tang2022swinunetr}
Tang, Y., Yang, D., Li, W., Roth, H.R., Landman, B., Xu, D., Nath, V., Hatamizadeh, A.:
Self-supervised pre-training of Swin transformers for 3D medical image analysis.
In: CVPR. pp. 20730--20740 (2022)

\bibitem{bar2022performance}
Bar, A., Gandelsman, Y., Darrell, T., Globerson, A., Efros, A.A.:
Performance prediction for semantic segmentation by a self-supervised image reconstruction decoder.
In: CVPR Workshops. pp. 4394--4403 (2022)

\bibitem{arthur2007kmeans}
Arthur, D., Vassilvitskii, S.:
k-means++: The Advantages of Careful Seeding.
In: Proc. ACM-SIAM Symposium on Discrete Algorithms (SODA) (2007)

\bibitem{conover1999practical}
Conover, W.J.:
Practical Nonparametric Statistics. 3rd edn.
John Wiley \& Sons (1999)

\bibitem{kendall2017uncertainties}
Kendall, A., Gal, Y.:
What uncertainties do we need in Bayesian deep learning for computer vision?
In: Advances in Neural Information Processing Systems (NeurIPS). pp. 5574--5584 (2017)

\bibitem{wright2021ranger21}
Wright, L., Demeure, N.:
Ranger21: A synergistic deep learning optimizer.
arXiv preprint arXiv:2106.13731 (2021)

\bibitem{paszke2019pytorch}
Paszke, A., Gross, S., Massa, F., et al.:
PyTorch: An Imperative Style, High-Performance Deep Learning Library.
In: Advances in Neural Information Processing Systems (NeurIPS) (2019)

\bibitem{cardoso2022monai}
Cardoso, M.J., Li, W., Brown, R., et al.:
MONAI: An Open-Source Framework for Deep Learning in Healthcare.
arXiv preprint arXiv:2211.02701 (2022)

\bibitem{maier2024metrics}
Maier-Hein, L., Reinke, A., Godau, P., et al.:
Metrics reloaded: Recommendations for image analysis validation.
Nature Methods \textbf{21}, 195--212 (2024)

\bibitem{jin2022oneshot}
Jin, Q., Yuan, M., Qiao, Q., Song, Z.:
One-shot active learning for image segmentation via contrastive learning and diversity-based sampling.
Knowledge-Based Systems \textbf{241}, 108278 (2022)

\bibitem{levy2026coldstart}
Levy, D., Assayag, B., Gaspar, L., Shimshoni, I., Specktor-Fadida, B.:
From Cold Start to Active Learning: Embedding-Based Scan Selection for Medical Image Segmentation.
arXiv preprint arXiv:2601.18532 (2026)

\bibitem{zhu2025medcalbench}
Zhu, N., Ma, X., Zhang, S., Wang, G.:
MedCAL-Bench: A Comprehensive Benchmark on Cold-Start Active Learning with Foundation Models for Medical Image Analysis.
arXiv preprint arXiv:2508.03441 (2025)

\end{thebibliography}

\end{document}